\documentclass[aps,prl,reprint,showpacs,floatfix,twocolumn]{revtex4-1}
\usepackage{graphicx}
\usepackage{textcomp}

\begin{document}

\title{Heralded Photonic Interaction between Distant Single Ions}

\author{M. Schug$^{1}$}
\email{mschug@physik.uni-saarland.de}
\author{J. Huwer$^{1,2}$}
\author{C. Kurz$^{1}$}
\author{P. M\"uller$^{1}$}
\author{J. Eschner$^{1}$}

\affiliation{
$^1$Universit\"at des Saarlandes, Experimentalphysik, Campus E2 6, 66123 Saarbr\"ucken, Germany\\
$^2$ICFO -- The Institute of Photonic Sciences, Avenida Carl Friedrich Gauss 3, 08860
Castelldefels (Barcelona), Spain}

\date{\today}

\begin{abstract}
We establish a heralded interaction between two remotely trapped single $^{40}$Ca$^+$ ions
through the exchange of single photons. In the sender ion, we release single photons with a 
controlled temporal shape on the P$_{3/2}$ to D$_{5/2}$ transition and transmit them to
the distant receiver ion. Individual absorption events in the receiver ion are detected
by quantum jumps. For continuously generated photons, the absorption reduces
significantly the lifetime of the long-lived D$_{5/2}$ state. For triggered single-photon
transmission, we observe a coincidence between the emission at the sender and quantum jump
events at the receiver.
\end{abstract}

\pacs{42.50.Ex, 42.50.Ct, 03.67.Hk, 42.50.Dv}

\maketitle

The realization of quantum networks requires the controlled transfer of quantum states
between quantum systems in a reversible manner, allowing for the distribution of
entanglement across the network \cite{Kimble2008}. One proposed experimental
implementation of such a network consists of single atoms as nodes, where quantum
information is stored and processed, and single photons for direct transmission between
the nodes \cite{Cirac1997}. In this context, it is a main prerequisite to control both
the emission and absorption of single photons by spatially separated single atoms or
atom-like systems.

For controlled single-photon emission, various systems exist \cite{Grangier2004} such as
neutral atoms \cite{Kuhn2002, McKeever2004}, trapped ions \cite{Keller2004, Barros2009},
or solid state systems \cite{Michler2000, Beveratos2002}, which
offer tailoring of specific properties of the emitted photons. Beyond that, entanglement
between a quantum system and its emitted photons \cite{Blinov2004, Volz2006, Wilk2007,
Stute2012} paves the way to transferring quantum information between distant network
nodes. These nodes may serve as bi-directional quantum interfaces, provided that also
control over the absorption of single photons by a single atom is attained
\cite{Cirac1997}. In this context, we have demonstrated the heralded photon absorption by
a single ion \cite{Piro2011, Huwer2012}. Controlling both the emission and absorption is
approached in recent experiments with two single quantum emitters such as organic
molecules \cite{Rezus2012} or atoms in optical cavities \cite{Specht2011, Ritter2012}.
Scalability of quantum networks to long distance quantum communication is enabled by
heralded entanglement between widely spaced atoms \cite{Moehring2007, Hofmann2012}, and
exploiting the quantum repeater scheme \cite{Briegel1998}. Additionally, hybrid
systems between different physical systems, such as entangled-photon sources and ions
\cite{Piro2011, Huwer2012}, entangled photons and quantum dots \cite{Polyakov2011}, or
atoms and ions \cite{Zipkes2011}, may be deployed to create building blocks of quantum
repeater systems.

Single trapped ions are ideal nodes for quantum networks, being particularly well
controlled in their motional and internal degrees of freedom and thereby allowing for
coherent processing of quantum information. Communication by photon transmission between
distant ions requires creation of single photons with controlled temporal shape and
coherence time \cite{Almendros2009}, ideally in a pure quantum state \cite{Kurz2012,
Stute2012}; at the same time, absorption of single photons must happen in a deterministic
or heralded manner \cite{Piro2011, Huwer2012, Kurz2012} in order to map the photonic
state to the ion. Such individual discrimination of absorption events has not been
implied in previous experiments \cite{Rezus2012, Specht2011, Ritter2012}.

In this letter we report the heralded interaction between distant single ions by emission and absorption of
individual photons. We perform photon transmission in continuous and sequence mode. Most
importantly, individual absorption events are detected by quantum jumps, and coincidence
between the photon generation and the absorption event is demonstrated.

Fig.~1a shows the double-trap apparatus consisting of two linear Paul traps separated by
approximately one meter. In each trap a single $^{40}$Ca$^{+}$ ion is confined and
excited by various laser beams for cooling and state manipulation. Along one of the axes,
two in-vacuum high numerical aperture laser objectives (HALOs) provide efficient optical
access to the ions \cite{Gerber2009}. In the sender ion trap we generate single 854~nm
photons of which we collect about $4\%$ with one HALO. We couple the photons into a
single-mode fiber that connects the two traps. In the receiver ion trap we use one HALO
to couple the 854~nm photons to the ion, and the other one to collect 397~nm fluorescence
photons which are detected with a photomultiplier tube (PMT).

The laser excitation scheme of the sender ion is illustrated in Fig.~1b. For triggered
generation of single 854~nm photons, the sequence starts with Doppler cooling using laser
light at 397~nm and 866~nm. Subsequently, an 850~nm laser couples the D$_{3/2}$ to the
P$_{3/2}$ level, thereby optically pumping the ion to the D$_{5/2}$ level and triggering
the emission of a single 854~nm photon. The beginning of the 850~nm laser pulse is used
as a time stamp, i.e.\ as a herald, for the photon generation. Finally, an 854~nm laser
repumps the ion to the S$_{1/2}$ level and concludes the sequence. For continuous
generation of 854~nm photons, all four lasers are constantly switched on.

\begin{figure}[htbp]
\centering
\includegraphics[width=0.45\textwidth]{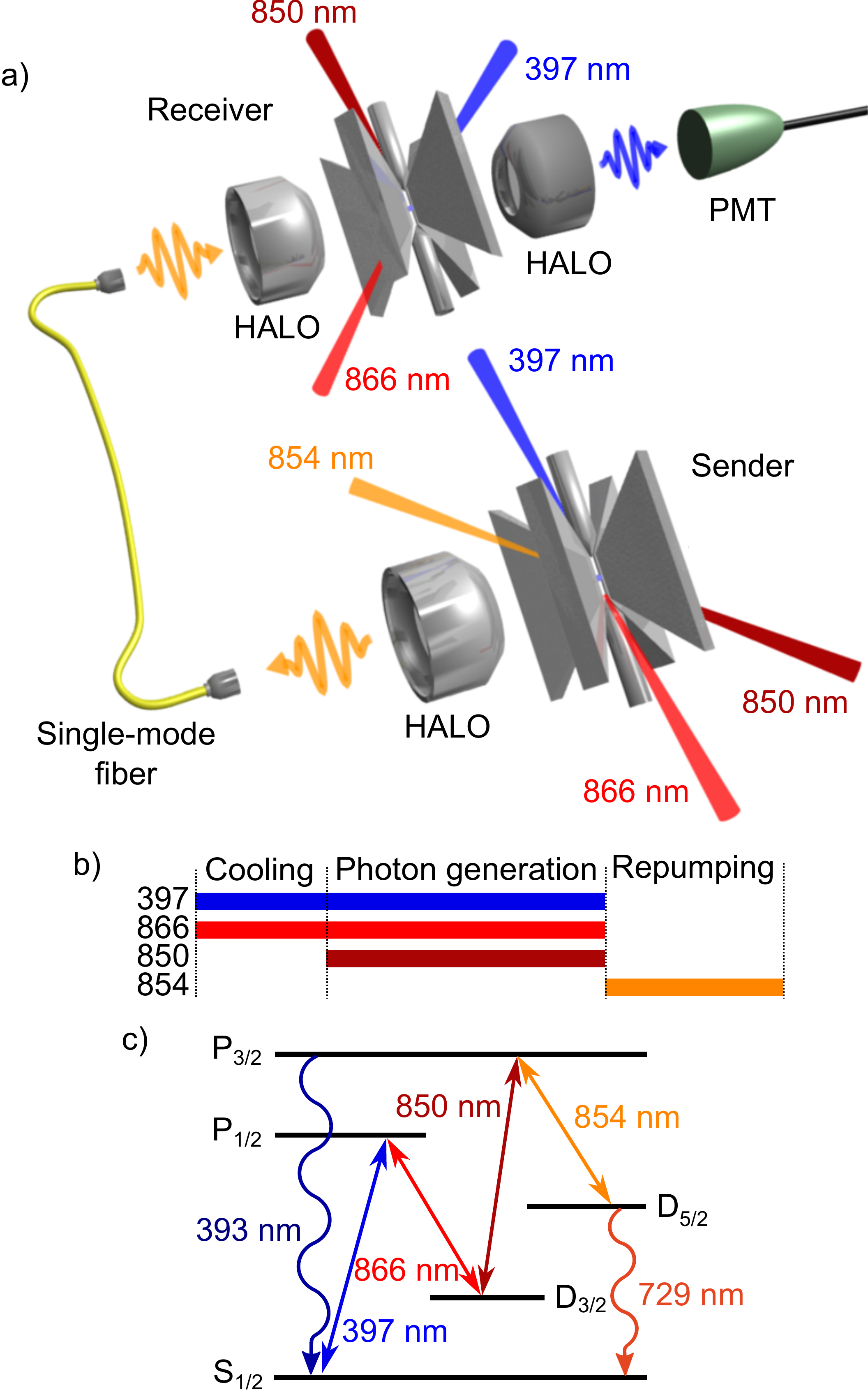}
\caption{(Color online) a) Schematic of the apparatus consisting of two linear Paul traps
separated by one meter distance and connected by a single-mode fiber. b) Laser excitation
scheme for triggered single-photon generation at the sender ion; see text for more
details. c) Level scheme and relevant transitions of the $^{40}$Ca$^+$ ion. The P$_{3/2}$
level decays with $2\pi\cdot 21.49$~MHz to S$_{1/2}$ and with $2\pi\cdot 1.35$~MHz to
D$_{5/2}$ \cite{Gerritsma2008}; the lifetime of the D$_{5/2}$ level is 1.1~s.}
\end{figure}

At the receiver ion, continuous-wave (cw) excitation with three lasers initializes the
absorption of 854~nm photons: the 397~nm and 866~nm lasers Doppler cool the ion, while
the emitted 397~nm fluorescence is monitored, and an attenuated 850~nm laser optically
pumps the ion to the metastable D$_{5/2}$ level. A quantum jump to that level is marked
by a sudden drop of the 397~nm fluorescence signal to the dark-count rate (a
bright-to-dark quantum jump). In the D$_{5/2}$ state the ion can absorb a photon from the
sender, which excites it to P$_{3/2}$. Subsequent decay to S$_{1/2}$ (with 93.47$\%$
probability \cite{Gerritsma2008}) is accompanied by the onset of the fluorescence (a
dark-to-bright quantum jump) which thus signals the absorption. Such a quantum jump
may also be caused by a spontaneous decay from D$_{5/2}$ to S$_{1/2}$. In the
following, this quantum jump scheme is used to observe and characterize the interaction
of the two ions via triggered and continuous transmission of photons.

Since we aim at high-rate interaction between the ions, we first optimize the efficiency
of the single photon generation at the sender ion, i.e.\ of the optical pumping process
into the D$_{5/2}$ level. We find that by employing a three-photon resonance condition
for the three lasers that connect S$_{1/2}$ and P$_{3/2}$ (397~nm, 866~nm, and 850~nm,
see Fig.~1c), the pumping rate is significantly increased over the one that would be
achieved by consecutive optical pumping (corresponding to using only the 397~nm and
850~nm lasers). This is corroborated by comparing the rates of scattered photons at
393~nm and 397~nm, $R_{393}$ and $R_{397}$, during the pumping: consecutive optical
pumping would yield $R_{393}/R_{397}=0.065$, equal to the branching ratio of P$_{1/2}$
into D$_{3/2}$ (after correcting for different detection efficiencies at the two
wavelengths). Our optimized scheme yields, in cw excitation conditions,
$R_{393}/R_{397}=1.34(2)$, hence the coherent three-photon process enhances the photon
generation rate by a factor of 20. Using such optimized excitation parameters, we measure
$4.5\cdot10^3$~s$^{-1}$ fiber-coupled 854 nm photons using an avalanche photodiode (APD) with
24(5)\% detection efficiency. From this we derive a continuous generation rate
$R_{854,\rm{cw}} = 18(4)$~kHz of 854~nm photons in a single optical mode.

In order to further characterize the single photons for the sender-receiver interaction,
we excite the sender ion to generate triggered single photons in sequence mode. The
photons are collected in a single-mode fiber, and their arrival times on an
APD are recorded with respect to the onset of the 850~nm laser pulse that triggers the
emission. Fig.~2 displays the arrival time distributions, i.e.\ the shapes of the
respective single-photon wave packets, for various power values of the 850~nm laser.
Variation of the 850~nm laser power allows us to control the wave packet duration $T_1$
(defined as the 1/e arrival time) between 0.83(3)~$\mu$s and 5.8(5)~$\mu$s
\cite{endnote_min_duration}.
We note that in contrast to the work of Refs.~\cite{Almendros2009, Kurz2012}, the photon
wave packets of Fig.~2 are incoherently broadened, i.e.\ the photons are far from
Fourier-limited, because of the unfavorable branching ratio of the P$_{3/2}$ level.
For the shortest measured photons, and for 125~kHz
repetition rate of the pulse sequence, the highest generation rate into a single spatial
(fiber) mode is $R_{854,\rm{triggered}}=3.0(6)$~kHz.

\begin{figure}[htbp]
\centering
\includegraphics[width=0.45\textwidth]{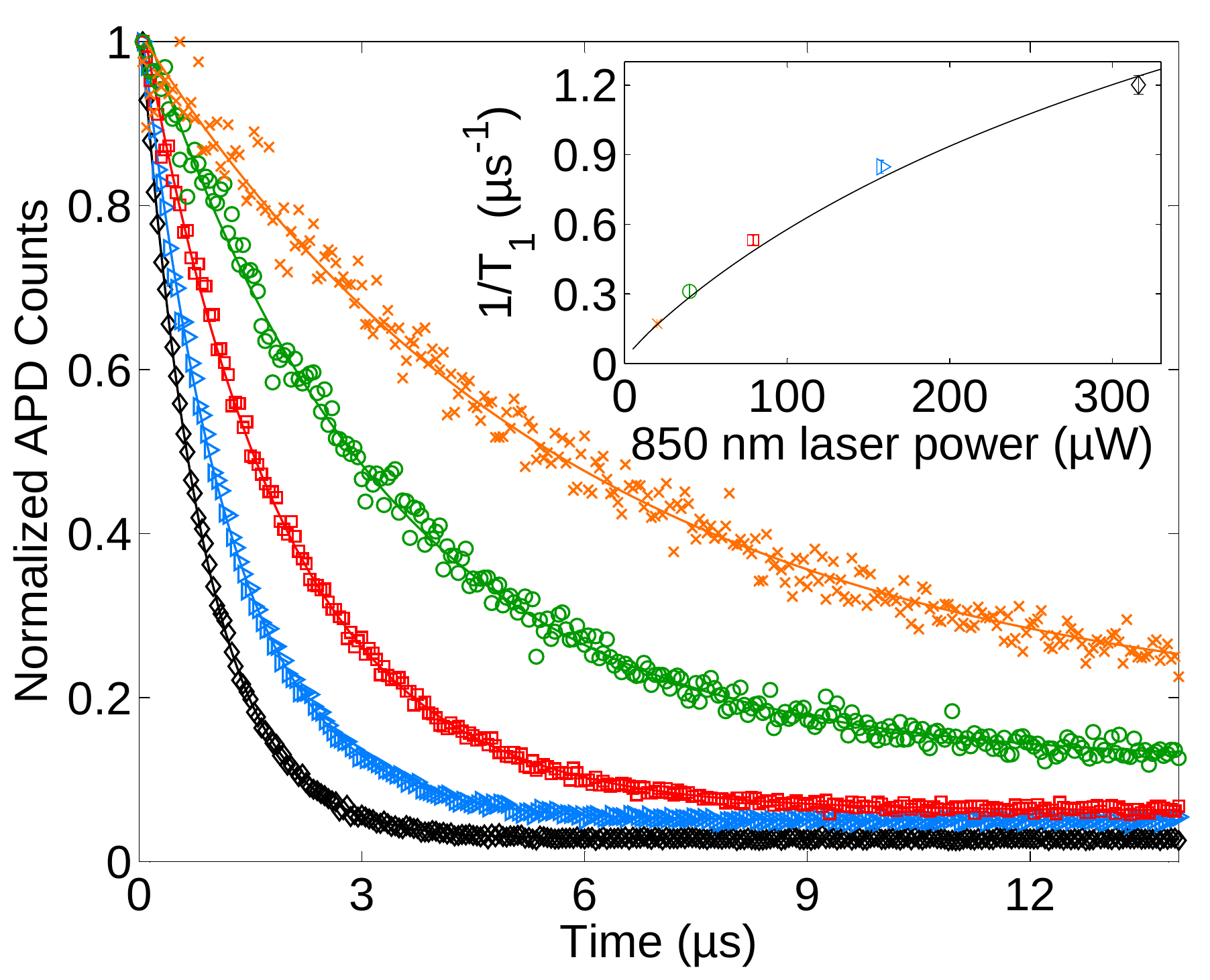}
\caption{(Color online) Distribution of single-photon ar\-ri\-val times for various values
of 850~nm laser power. Time zero is marked by switching on the 850~nm laser beam.
Experimental data are displayed with 50~ns time resolution. The lines are exponential
fits to the data from which the photon wave packet durations $T_1 = \{0.833(3), 1.18(4),
1.88(7), 3.22(27), 5.8(5) \}$~$\mu$s are deduced. Each curve corresponds to 15~min of
acquisition at a repetition rate of 55.5~kHz (first three curves) resp.\ 30~kHz (last two
curves). The inset shows the inverse wave packet duration 1/$T_1$ as a function of the
850~nm laser power. The solid line is calculated by numerically solving the optical Bloch
equations including all Zeeman sublevels.}
\end{figure}

A first cw interaction measurement is performed using continuous generation of 854~nm
photons at the sender ion while monitoring quantum jumps at the receiver ion. It is
illustrated by Fig.~3. When the 850~nm laser at the sender is switched off, such that no
photons are generated, dark-to-bright fluorescence jumps at the receiver ion mark
spontaneous decay of the D$_{5/2}$ level. From a 20~min fluorescence trace we extract
the effective lifetime of the D$_{5/2}$ level given by the average length of the dark period
$\tau_{\rm{off}}=1022(33)$~ms, and the respective decay rate
$R_{\rm{off}}=0.97(3)$~s$^{-1}$ (see Fig.~3a). The value of $\tau_{\rm{off}}$ is slightly
lower than the literature value of 1168(9)~ms \cite{Kreuter2005}; this may be caused by
854~nm background light entering the trap.

\begin{figure}[hbt]
\includegraphics[width=0.4\textwidth]{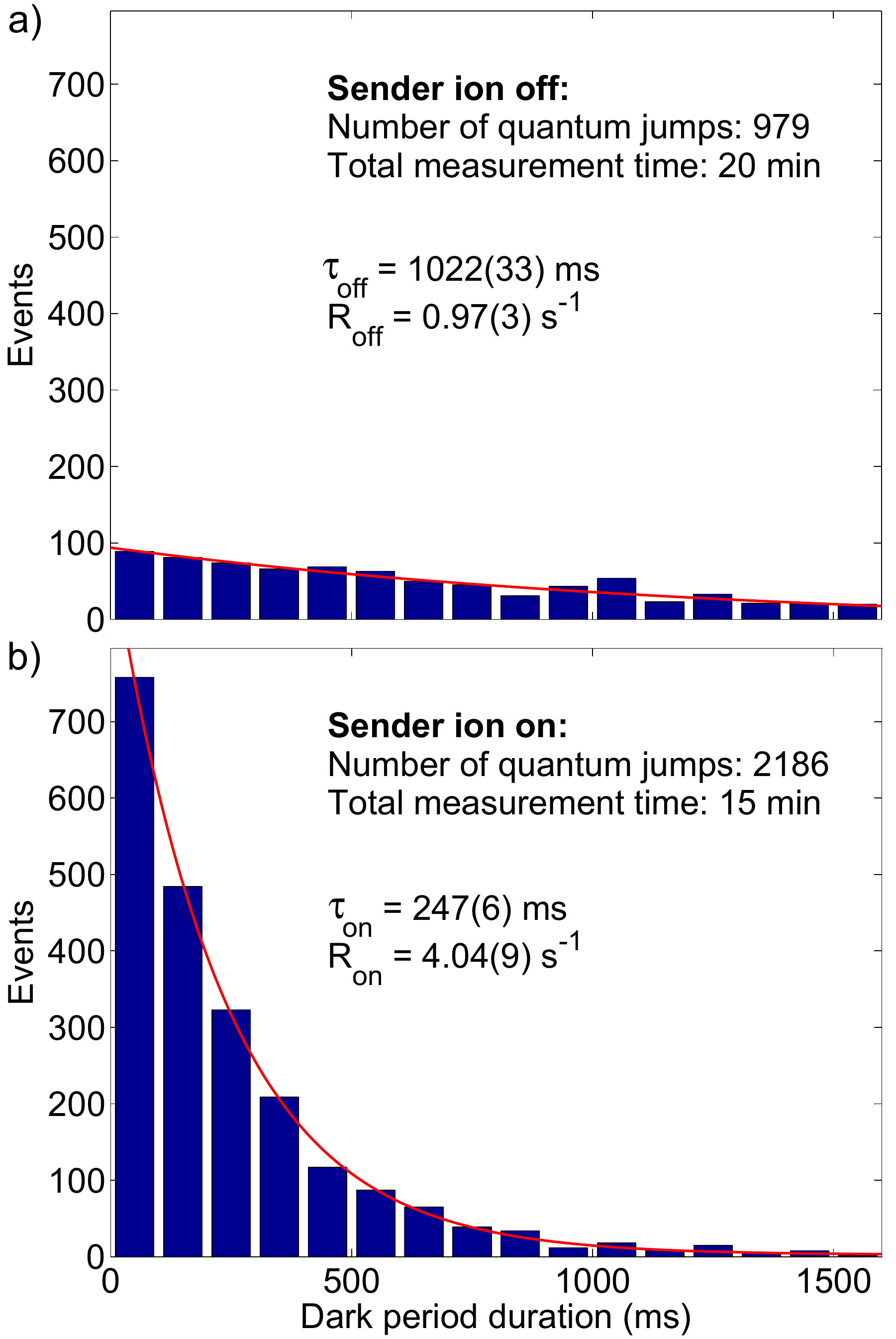}
\caption{(Color online) Histogram of dark period durations in the receiver ion after a
quantum jump to D$_{5/2}$; a) without photons from the sender ion, and b) with photons
continuously transmitted. The mean values of the dark periods $\tau_{\rm{on,off}}$ and
the corresponding decay rates $R_{\rm{on,off}}=1/\tau_{\rm{on,off}}$ are also displayed.}
\end{figure}

When the 850~nm laser at the sender ion is switched on, a significant reduction of the
dark period duration to $\tau_{\rm{on}}=247(6)$~ms is observed, corresponding to a decay
rate $R_{\rm{on}}=4.0(1)$~s$^{-1}$ (see Fig.~3b). From the two values, we find the
absorption rate of 854~nm photons, $R_{\rm{abs}} = R_{\rm{on}}-R_{\rm{off}} =
3.0(1)$~s$^{-1}$. This rate was measured at about 12~kHz photon transmission rate, hence
we conclude that a single photon is absorbed by the receiver ion with
$P_{\rm{abs}}=2.5(5)\cdot10^{-4}$ probability.

We obtain an independent value for the single-photon absorption probability at the
receiver ion by sending photons from a 854 nm cw laser through the same fiber that we use
for transmitting the single photons from the sender ion. The laser is attenuated to an
intensity that results in about the same APD count rate as for continuous single-photon
generation. Under these conditions we find an absorption probability of
$P_{\rm{laser}}=4.3(9)\cdot10^{-4}$. We attribute the reduction in the case of
single-photon transmission to frequency broadening of the sender photons: while the
single-frequency laser is tuned to the peak of the absorption spectrum, the generated
single photons have various frequency components, corresponding to the involved Zeeman
transitions; additionally, their rapid generation leads to a spectral width of the
individual photons of about 6~MHz. Numerical calculations taking these broadening
mechanisms into account confirm the measured reduction within the error margins.

\begin{figure}[hbt]
\includegraphics[width=0.45\textwidth]{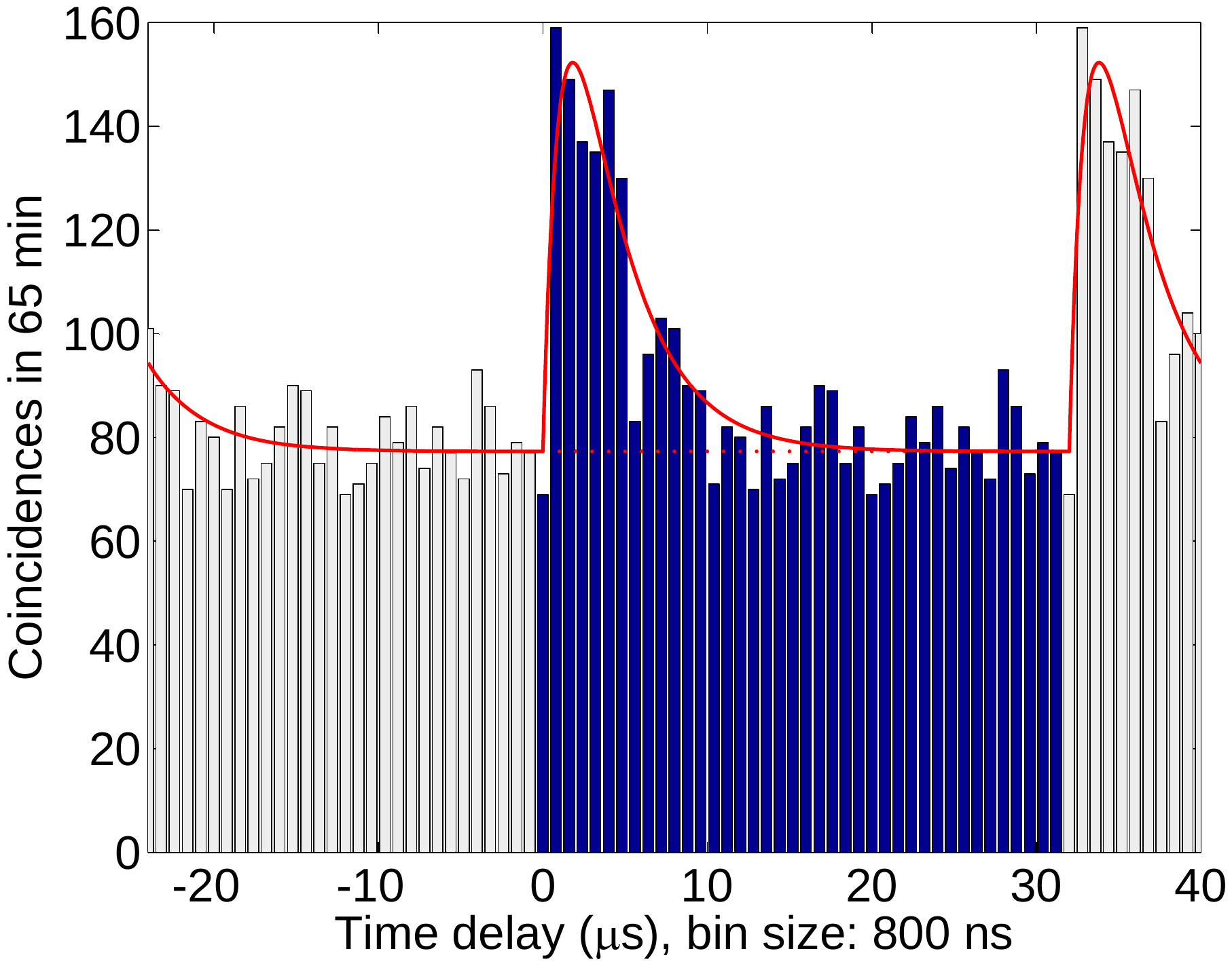}
\caption{(Color online) Temporal correlation function (time delay distribution) between
single-photon emission triggers at the sender ion and absorption events (quantum jumps)
at the receiver ion. These data were recorded with 31.25~kHz sequence repetition rate for
a total acquisition time of 65~min, and are displayed with 0.8~$\mu$s bin size. The red
curve shows the expected dependence, and the dotted line indicates the background level
(details see text). The signal repeats with the sequence repetition time (grey areas).}

\end{figure}

Finally, we transmit triggered single photons from sender to receiver, operating the
sender ion in sequence mode. In order to demonstrate that their absorption is signaled by
a dark-to-bright quantum jump, we temporally correlate the emission trigger at the sender
with the first 397~nm fluorescence photon that we detect at the receiver after a quantum
jump. The data are shown in Fig.~4.
Trigger-absorption coincidence is evidenced by a peak in the correlation function on top
of a background of uncorrelated events. Starting from zero time delay, a steep rise is
visible from 0.8~$\mu$s to 1.6~$\mu$s, reflecting the photon wave packet duration that
was set to $\tau_{854}=1.1~\mu$s. The rise is followed by an exponential decay with
3~$\mu$s time constant, corresponding to the average time between absorption and
detection of the first photon, i.e.\ to the inverse of the detection rate of 397~nm
photons, $R_{397}=3\cdot10^5$~Hz.
The convolution of these two exponentials (red curve in Fig.~4) fits the data very well.
The background (dotted line in Fig.~4) of $\sim$80 events per 0.8~$\mu$s time bin arises
from uncorrelated pairs of emission triggers and quantum jumps induced by spontaneous
decay of the D$_{5/2}$ level.

In summary, we generate single-mode 854~nm photons with controlled temporal shape from a
single $^{40}$Ca$^+$ ion through resonant three-photon excitation, and we demonstrate
their absorption, signaled by a quantum jump, in another ion at $\sim 1$~m distance. In
continuous mode, up to $1.8\cdot10^4$ photons/s are transmitted between the ions, and
single-photon absorption results in a reduction to 24\% of the D$_{5/2}$ lifetime at the
receiver ion. In sequence mode, the onset of a trigger laser heralds the creation of a
transmitted photon, and the correlation function between the emission trigger and the
photon absorption signal reveals the coincidence of these processes within the time
resolution of about 3~$\mu$s, set by the detection rate of fluorescence photons. With
125~kHz sequence repetition frequency, the photon transmission rate is 3~kHz, and about
0.025\% of these photons are absorbed in the receiver ion. Taking the emission trigger as
a herald for the absorption events, the overall heralding efficiency is 6$\cdot10^{-6}$.

Our experiment is a proof of principle of heralded individual photon absorption in a
single atom. In order to preserve the photonic state in the atom and thereby realize a
heralded quantum memory, we will combine our method with the detection of the
Raman-scattered photon created in the absorption process, which we recently demonstrated
with 1.5\% efficiency \cite{Kurz2012}. This will also make the heralding independent of
the emission trigger, thereby allowing for, e.g., entanglement transfer between photon
and atom pairs \cite{Sangouard2013}. Among the systems with which single-photon
absorption by a single absorber has been observed \cite{Rezus2012, Specht2011,
Ritter2012}, single ions stand out for their unique prospect of integrating atom-photon
interfaces with multi-qubit quantum logic. The results also offer the perspective of
photonic interaction between different single quantum systems in a hybrid quantum
network, e.g.\ between ions and solid-state emitters or absorbers \cite{Neu2012},
possibly assisted by quantum frequency conversion \cite{Zaske2012} to bridge the gap
between different wavelengths.

~\\

\begin{acknowledgments}
We acknowledge support by the BMBF (Verbundprojekt QuOReP, CHIST-ERA project QScale), the German Scholars Organization/Alfried Krupp von Bohlen und Halbach-Stiftung, the EU (AQUTE Integrating Project) and the ESF (IOTA COST Action).
\end{acknowledgments}

{}

\end{document}